\documentclass[aps,prb,twocolumn,superscriptaddress,notitlepage]{revtex4-1}
\usepackage{graphicx}
\usepackage[caption=false]{subfig}
\usepackage{float}
\usepackage{natbib}
\usepackage{xcolor}
\usepackage{amsmath}
\usepackage{amssymb}
\usepackage{array}
\usepackage{wasysym}
\usepackage{color,soul}
\usepackage{braket}
\usepackage{verbatim}
\usepackage{hyperref}
\usepackage{gensymb}
\usepackage{url}
\newcommand{\bmp}{BaMo(PO$_4$)$_2~$}
\newcommand{\bmpns}{BaMo(PO$_4$)$_2$}
\usepackage{filecontents}
\begin{document}

\title{Large Easy Axis Anisotropy in the One-Dimensional Magnet \bmp}

%\author{Author list goes here...}
\author{Aly H. Abdeldaim}
\email[Email address:~]{aly.abdeldaim@liverpool.ac.uk}
\affiliation{Department of Chemistry and Chemical Engineering, Energy \& Materials, Chalmers University of Technology, SE-412 96 Gothenburg, Sweden}
\affiliation{Department of Chemistry and Materials Innovation Factory, University of Liverpool, 51 Oxford Street, Liverpool, L7 3NY, UK}
\author{Danis I. Badrtdinov}
\affiliation{Theoretical Physics and Applied Mathematics Department, Ural Federal University, 620002 Yekaterinburg, Russia}
\author{Alexandra S. Gibbs}
\affiliation{ISIS Neutron and Muon Source, Science and Technology Facilities Council, Didcot OX11 0QX, United Kingdom}
\author{Pascal Manuel}
\affiliation{ISIS Neutron and Muon Source, Science and Technology Facilities Council, Didcot OX11 0QX, United Kingdom}
\author{Helen C. Walker}
\affiliation{ISIS Neutron and Muon Source, Science and Technology Facilities Council, Didcot OX11 0QX, United Kingdom}
\author{Manh Duc Le}
\affiliation{ISIS Neutron and Muon Source, Science and Technology Facilities Council, Didcot OX11 0QX, United Kingdom}
\author{Chien Hung Wu}
\affiliation{ISIS Neutron and Muon Source, Science and Technology Facilities Council, Didcot OX11 0QX, United Kingdom}
\author{Dariusz Wardecki}
\affiliation{Department of Chemistry and Chemical Engineering, Energy \& Materials, Chalmers University of Technology, SE-412 96 Gothenburg, Sweden}
\author{Sten-Gunnar Eriksson}
\thanks{Deceased}
\affiliation{Department of Chemistry and Chemical Engineering, Energy \& Materials, Chalmers University of Technology, SE-412 96 Gothenburg, Sweden}
\author{Yaroslav O. Kvashnin}
\affiliation{Department of Physics and Astronomy, Uppsala University, P.O. Box 516, S-75120 Uppsala, Sweden}
\author{Alexander A. Tsirlin}
\email{Email address: altsirlin@gmail.com}
\affiliation{Theoretical Physics and Applied Mathematics Department, Ural Federal University, 620002 Yekaterinburg, Russia}
\affiliation{Experimental Physics VI, Center for Electronic Correlations and Magnetism, Institute of Physics, University of Augsburg, 86135 Augsburg, Germany}
\author{G\o{}ran J. Nilsen}
\email[Email address:~]{goran.nilsen@stfc.ac.uk}
\affiliation{ISIS Neutron and Muon Source, Science and Technology Facilities Council, Didcot OX11 0QX, United Kingdom}
\date{\today}

\begin{abstract}
We present an extensive experimental and theoretical study on the low-temperature magnetic properties of the monoclinic anhydrous alum compound \bmpns. The magnetic susceptibility reveals strong antiferromagnetic interactions $\theta_{CW} = -167$~K and long-range magnetic order at $T_N=22$~K, in agreement with a recent report. Powder neutron diffraction furthermore shows that the order is collinear, with the moments near the $ac$ plane. Neutron spectroscopy reveals a large excitation gap $\Delta = 15$~meV in the low-temperature ordered phase, suggesting a much larger easy-axis spin anisotropy than anticipated. However, the large anisotropy justifies the relatively high ordered moment, N\'{e}el temperature, and collinear order observed experimentally, and is furthermore reproduced in a first principles calculations using a new computational scheme. We therefore propose \bmp to host $S=1$ antiferromagnetic chains with large easy-axis anisotropy, which has been theoretically predicted to realize novel excitation continua.
\end{abstract}

\maketitle

\section{INTRODUCTION}
\begin{figure}
\begin{center}
{\resizebox{0.87\columnwidth}{!}{\includegraphics{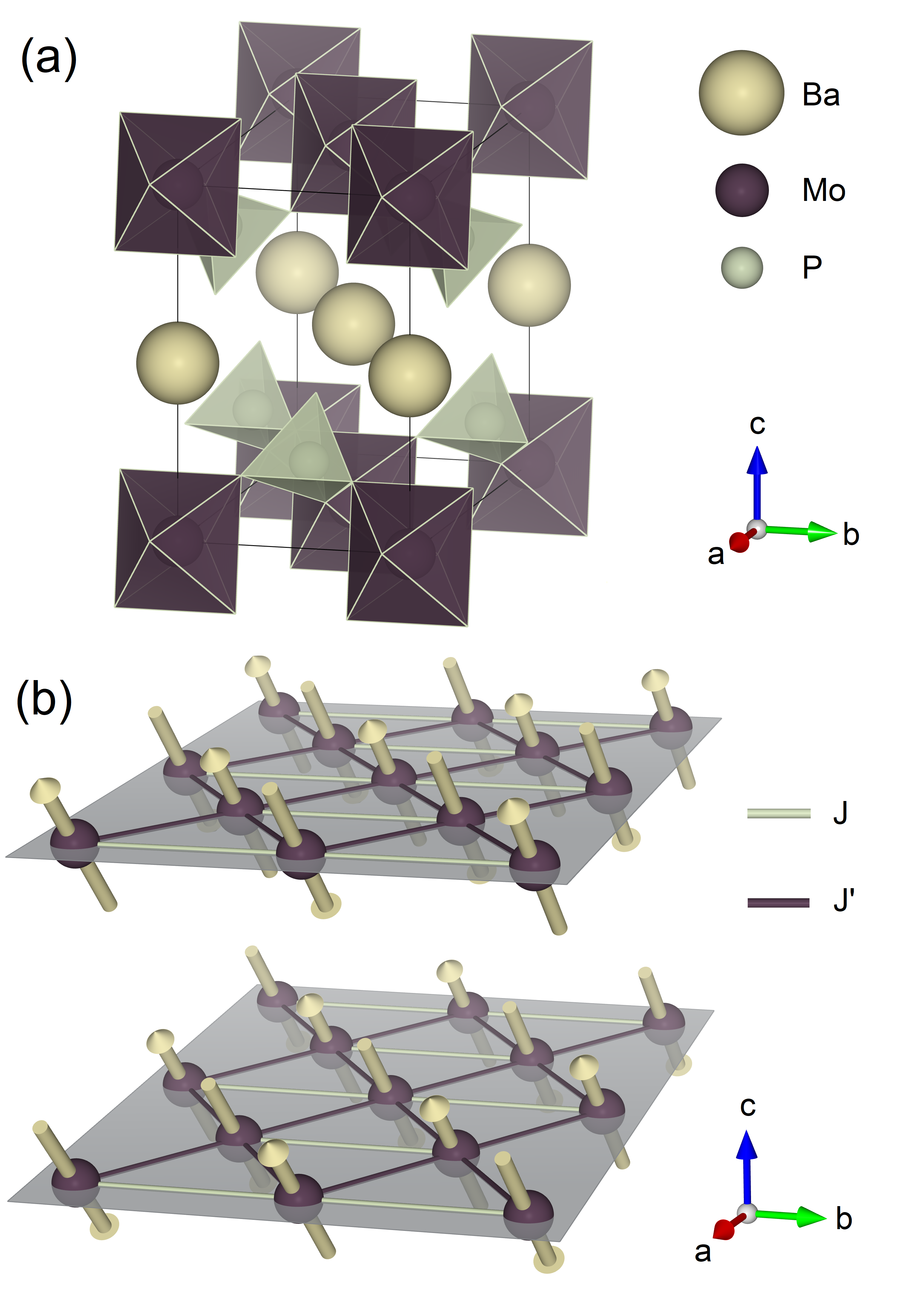}}}
%\hspace*{+1.3cm} (a)
\end{center}
\caption{(a) The crystal structure of BaMo(PO$_4$)$_2$ as observed along the $\langle 111\rangle$ direction. Here, Mo$^{+4}$ is presented in purple, Ba$^{2+}$ in yellow, and P$^{5+}$ in green. (b) An illustration of the anisotropic triangular arrangement of Mo$^{4+}$ moments and their suggested directions in the magnetic structure of \bmp based on the refinement of NPD data at 1.5 K where the intra- (J) and inter-chain (J$^\prime$) exchange interactions are depicted in green and purple, respectively. The structure was generated using the VESTA visualization software \cite{Vesta}.}\label{fig:BMP_nuclear}
\end{figure}

Low dimensionality, geometric frustration, and anisotropic interactions are all aspects which promote novel quantum phenomena in magnetic systems. To give just a few examples, models like the $S=1/2$ kagome lattice antiferromagnet (low-dimensional, frustrated) \cite{Kagome}, the Kitaev honeycomb model (low-dimensional, anisotropic) \cite{KITAEV20062,Jackeli}, and the pyrochlore lattice antiferromagnet with exchange anisotropy (frustrated, anisotropic) \cite{pyrochlore}, support a range of exotic spin liquid ground states, often with fractionalized quasi-particle excitations. Theoretical interest in these models has driven an extensive search for experimental realizations, and more broadly, for new materials which contain the above ingredients in new and unexpected combinations. In this context, a nearly untapped source of candidate materials are the anhydrous alum family $AB$($CX_4$)$_2$, where $A$ is an alkali metal cation, $B$ can be a range of magnetic transition metal, lanthanide, or actinide cations, and $CX_4$ is a polyatomic anion like sulfate (SO$_4$)$^{2-}$ or phosphate (PO$_4$)$^{3-}$. Depending on the relative $A$ and $B$-site ionic radii, the anhydrous alums may crystallize in either rhombohedral, trigonal, or monoclinic structures. In the monoclinic case, the $B$-site magnetic ions form an anisotropic triangular lattice as illustrated in Fig.\ref{fig:BMP_nuclear}, whilst in the other two cases, an isotropic (equilateral) triangular lattice is realised \cite{BREGIROUX201526}. Given this, as well as their chemical flexibility, a range of low-dimensional, frustrated, and anisotropic models can potentially be realized in the anhydrous alums.

Among the relatively few previously studied monoclinic anhydrous alums, most realize the one-dimensional limit of the anisotropic triangular lattice model, \textit{i.e.} $J> J^\prime$ [Fig. \ref{fig:BMP_nuclear}]. For example, a recent investigation of the $S=1/2$ Heisenberg compound KTi(SO$_4$)$_2$ revealed a ratio between the inter-chain and intra-chain exchanges $J^\prime / J < 0.15$ \cite{KTS}, while the magnetic structure of the classical system KFe(SO$_4$)$_2$ appears to be compatible with $J^\prime /J = 0.25$ \cite{KFS}. Due to the frustrated interchain couplings in these materials, long range antiferromagnetic ordering is strongly suppressed, with the ratio of the Curie-Weiss constant to ordering temperature $\theta_{CW}/T_N$ typically exceeding $10$; indeed, KTi(SO$_4$)$_2$ does not appear to order down to $50$~mK despite a dominant antiferromagnetic exchange of $J = 15$~K, and exhibits fractionalized continua of spinon excitations. These features highlight the potential of the family to realize interesting  magnetic phenomena.

Because many monoclinic anhydrous alums are magnetically quasi one-dimensional, compounds with $S=1$ are of particular interest; the $S=1$ Heisenberg chain model famously has a topologically ordered \cite{denNijs, Kennedy} singlet ground state with a Haldane gap to the first excited triplet \cite{Haldane}. Both single- and multi-triplet excitations have been observed above the gap using inelastic neutron scattering on several materials \cite{miller2001magnetism}, such as CsNiCl$_3$ \cite{CsNiCl3}, Ni(C$_2$H$_8$N$_2$)$_2$NO$_2$ClO$_4$ \cite{Regnault}, and AgVP$_2$S$_6$\cite{AgVP2S6}. These experiments also highlight the role of terms beyond the nearest-neighbor Heisenberg exchange; most notably, the combination of spin-orbit coupling and crystal electric fields in the $d^2$ or $d^8$ electronic configurations which give rise to $S=1$ result in single-ion ($3d$, $4d$ and $5d$ ions) or exchange ($4d$ and $5d$) anisotropies that split the triplet excitations or even entirely suppress the Haldane gap. The generalized model Hamiltonian containing both single-ion and exchange anisotropy hosts three gapped phases separated by quantum critical points, depending on the type and relative magnitude of the anisotropy \cite{Chan}; in addition to the Haldane phase, a so-called large-$D$ state with $S^z=0$ on each site is favoured for dominant easy plane single-ion anisotropy, while a magnetically ordered N\'{e}el phase results from easy-axis single-ion anisotropy. Common to all of these states are the presence of excitation continua associated with the combination of quantum spins and low dimensionality; the large $D$ phase hosts continua which involve pairs of sites being excited from $S^z=0$ to $S^z=\pm 1$ \cite{Papanicolaou_1990}, while the N\'{e}el phase can be mapped onto an effective $S=1/2$ model with spinons carrying $S^z_{eff}=\pm 1/2$ and empty sites, or holons, with $S^z=0$ \cite{denNijs, Rommelse}. 

The effect on this picture of a frustrating interchain coupling, such as that resulting from the structure of the anhydrous alums, is not well known, although studies on the $S=1$ Heisenberg chain on the anisotropic triangular lattice suggest that the Haldane gap is only suppressed at $J^\prime /J \sim 0.4$ \cite{Trumper}, as compared to $J^\prime/J\sim 0.04$ for the unfrustrated case \cite{Matsumoto}. The magnetically ordered state that results upon the suppression of the gap in the frustrated (anisotropic triangular) case is expected to be incommensurate along the chain direction, as observed in the classical system KFe(SO$_4$)$_2$\cite{KFS}. We finally note that in the fully frustrated triangular case, \textit{i.e.} $J^\prime = J$, the presence of anisotropic exchange has been predicted to lead to at least two different types of spin liquid state \cite{Chernyshevr}.  

The present study focuses on the monoclinic anhydrous alum \bmpns, where the Mo$^{4+}$ ($d^2$) ions form a distorted anisotropic triangular lattice of $S=1$ spins. The inclusion of a magnetic $4d$ cation means that anisotropic terms, both single-ion and exchange, should be much more prominent than in the $3d$ systems mentioned above. Indeed, a previous experimental and theoretical study of \bmpns~ indicates that the system orders at a relatively high temperature $T_N = 21$~K relative to the dominant exchange of $J\sim40$~K, \textit{i.e} $\theta_{CW}/T_N\sim 2$\cite{danis2018}\textsuperscript{,}\footnote{Our study and that reported in Hembacher et al. were initiated independently and run in parallel until the publication of \cite{danis2018}. Despite the different sample preparation methods and qualities between the two studies, many of the conclusions drawn from the thermodynamic and neutron diffraction measurements (here reported in Sections~\ref{sec:exp}$-$\ref{sec:results}) are the same.}. Since the interchain coupling is estimated to be small $\sim 0.1J$, well below the threshold to close the Haldane gap, and because the magnetic structure is collinear, strong easy-axis anisotropy is anticipated. However, the single-ion anisotropy parameter predicted from previous electronic structure calculations, $D\sim 0.1J$, is also too small to close the Haldane gap, leaving the ordering mechanism in \bmp an open question. 

In this paper, we will show that the high magnetic ordering temperature and collinear ordered structure of \bmp can in fact be explained by a strong single-ion anisotropy. We will begin by reporting a new synthetic route to powder samples with fewer paramagnetic impurities and higher crystallinity than previous in Section II. In Section III, we will discuss our magnetic susceptibility, specific heat, and neutron diffraction data, making comparisons with previous results. Most notably, our new samples exhibit a peak in specific heat where none was seen before, and a larger Curie-Weiss constant $\theta = -167$~K. Then, in Section IV, we will provide evidence for the strong easy-axis anistropy via inelastic neutron scattering measurements of the spin excitation spectrum, which we find to exhibit a large anisotropy gap of $\Delta =15$~meV and a narrow bandwidth of $5$~meV. Linear spin wave fits to this data allow us to approach a determination of the full anisotropic Hamiltonian of the system; the best fit to the inelastic neutron scattering data suggests either a moderate easy-plane or easy-axis exchange anisotropy in addition to the strong single easy-axis single-ion anisotropy. The experimental single-ion anisotropy direction and magnitude are qualitatively reproduced by improved electronic structure calculations, as discussed in Section V. The former, in particular, is found to be exceptionally sensitive to the on-site Hund's coupling $J_H$, justifying the incorrect ordered moment direction found previously. We finally discuss the possibilities for observing excitation continua in inelastic neutron scattering studies in Section VI, before concluding in Section VII.

\section{EXPERIMENTAL METHODS}\label{sec:exp}
The preparation of the polycrystalline \bmp samples used in this study follows a distinct procedure from the previously reported synthetic routes \cite{leclaire,danis2018}. Stoichiometric amounts of BaCO$_3$ (Alfa Aesar 99.9\%), MoO$_2$ (Alfa Aesar 99.0\%), and (NH$_4$)$_2$HPO$_4$ (Alfa Aesar 98+\%) were intimately mixed and heated for 12 hours in a tube furnace at 873 K under vacuum. The reaction proceeded through a subsequent heat treatment at 1173 K for 48 hours followed by a final 72 hour heating stage at 1273 K. 

DC magnetic susceptibility measurements were carried out on a 52.7 mg sample using a Quantum Design MPMS SQUID magnetometer. Data were collected using both the zero-field-cooled (ZFC) and field-cooled (FC) protocols under a 1000 Oe applied field and over a 2 - 300 K temperature interval. Specific heat data were recorded at zero field in the temperature range 2 - 300 K using the thermal relaxation method on a Quantum Design PPMS.

The presence of \bmp and minor MoO$_2$ and Ba$_2$P$_2$O$_7$ impurities was confirmed using a Bruker D8 X-ray powder diffractometer with Cu $K \alpha$ radiation. Further structural characterization was performed using two neutron powder diffraction (NPD) instruments at the ISIS Neutron and Muon Source, UK. The low-temperature nuclear structure of \bmp was first verified using the time-of-flight (TOF) High-Resolution Powder Diffractometer (HRPD) \cite{HRPD}; the experiment was conducted on a 5 g sample in a flat plate vanadium-windowed container and the measurement temperature of $2$ K was reached using a standard $^4$He cryostat. Data collected from all three fixed angle detector banks (centered on 168\degree, 90\degree, and 30\degree) were used to determine the crystal structure. For the magnetic structural determination, additional NPD measurements were performed using the long-wavelength WISH TOF diffractometer \cite{WISH}. Data were recorded on 10 fixed angle detector banks in $5$ pairs at $1.5$~K$<T_N$ and $30$~K$>T_N$ using a $10$~g sample contained in a vanadium cylinder. The nuclear and magnetic structures were refined using the Rietveld method with the \texttt{GSAS}\cite{GSAS,GSAS2} and the \texttt{FullProf}\cite{Fullprof} software packages, respectively, with LeBail profile fitting being used for the impurity phases. Finally, the dynamical structure factor $S(Q,\Delta E = \hbar \omega)$ of \bmp was measured using the same 10 g powder sample on the MERLIN direct geometry TOF spectrometer, also at ISIS \cite{Merlin}. The sample was placed in a closed-cycle refrigerator, and spectra were collected using incident neutron energies of 50.4 meV, 19 meV and 10 meV at 5 K, 10 K, 20 K, 40K, and 60 K. 

\section{Results}\label{sec:results}
\subsection{Nuclear Structure}
\begin{figure}
\begin{center}
{\resizebox{0.995\columnwidth}{!}{\includegraphics{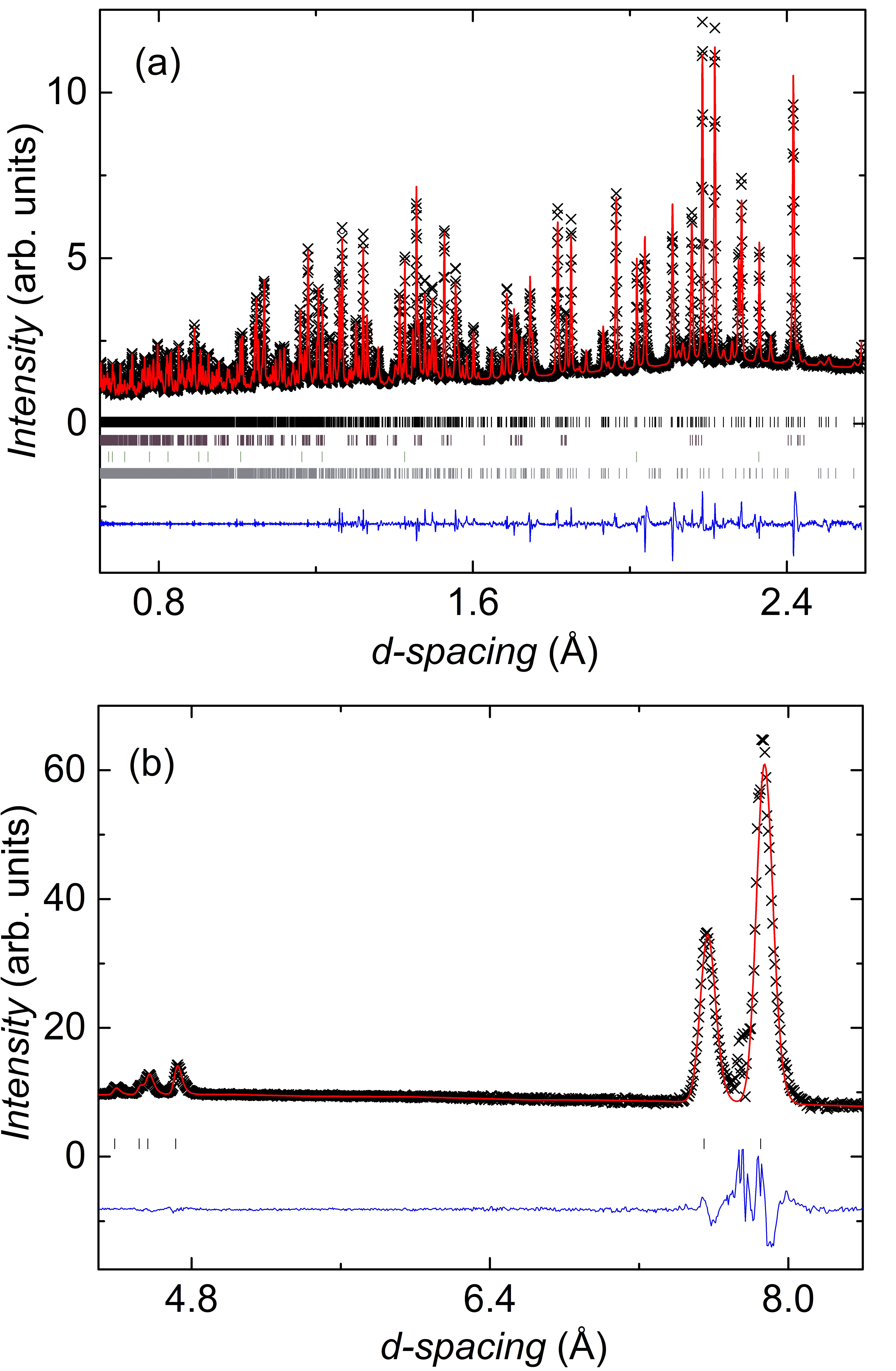}}}
%\hspace*{+1.3cm} 
%\hspace{+5.4cm} (a)
%\hspace{+9.2 cm} (b)
\end{center}
\caption{(a) Rietveld refinement of the \bmp nuclear structure according to data collected on the first bank of HRPD at 2 K. Observed and calculated points are shown in red and black, respectively, and their difference is shown in blue. The allowed Bragg reflections for all the refined phases are depicted in black (BaMo(PO$_4$)$_2$), purple (MoO$_2$), green (Al), and grey (Ba$_2$P$_2$O$_7$) tick marks. The agreement with the calculated pattern is good, with $R_{wp}=2.55\%$, $\chi^2=10.7$. (b): Rietveld refinement of the P$_s\overline{1}$ magnetic structure using 1.5 K - 30 K subtracted data collected on the WISH instrument. Observed data points are shown in red, the calculated pattern in black, their difference in blue, and the magnetic reflections are indicated with black tick marks.}\label{fig:HRPD}
\end{figure}

The analysis of the high-resolution NPD data from the HRPD instrument identified the monoclinic $C 2/m$ space group as describing the structure of \bmp at all measured temperatures. The lattice parameters at $2$~K are $a=8.1778(7)$ \AA, $b=5.2784(5)$ \AA, $c=7.8024(7)$ \AA, and $\beta=94.858(1)$\degree, with the remaining refinement [Fig.~\ref{fig:HRPD}(a)] parameters reported in Table.~\ref{table:HRPDdata}. The overall structural model is consistent with those reported in Leclaire \textit{et al.} \cite{leclaire} and Hembacher \textit{et al.} \cite{danis2018}. As depicted in Fig.~\ref{fig:BMP_nuclear}, the crystal structure consists of a quasi-two-dimensional layered motif of [Mo(PO$_4$)$_2$]$^{2-}$ sheets in the $ab$ plane, separated by Ba$^{2+}$ cations in the $c$ direction. The most relevant magnetic superexchange pathways are between intra-chain ($J$, 5.28 \AA, $J_2$ in \cite{danis2018}) and inter-chain neighboring ($J^\prime$, 4.86 \AA, $J_1$ in \cite{danis2018}) Mo$^{4+}$ cations \textit{via} the bridging (PO$_4$)$^{3-}$ groups. These pathways run along the $\langle$010$\rangle$ and $\langle$110$\rangle$ directions, respectively (Fig.~\ref{fig:BMP_nuclear}). \begin{table}[H]
\caption{Crystallographic information as obtained from the refinement of HRPD data at 2 K. The unit cell parameters are $a=8.1778(7)$ \AA, $b=5.2784(5)$ \AA, $c=7.8024(7)$ \AA, and $\beta=94.858(1)\degree$.}
\begin{ruledtabular}
\begin{tabular}{@{}lllll}
Atom & x & y & z & U\textsubscript{iso} (\AA\textsuperscript{2})  \\ \hline
Ba & 0  & 0 & 0 & 0.00400(38) \\
Mo & 0  & 0 & 0.50000  & 0.00621(32) \\
P & 0.13131(11) & 0.50000 & 0.29140(11) & 0.00426(28) \\
O1 & 0.02184(6) & 0.26299(9) & 0.31072(7) & 0.00339(21) \\
O2 & 0.26456(9) & 0.50000  & 0.43974(9) & 0.00352(21) \\
O3 & 0.18737(10) & 0.50000 & 0.11368(10) & 0.00566(21) \\
\end{tabular}
\end{ruledtabular}
\label{table:HRPDdata}
\end{table}

\subsection{Magnetic Susceptibility and Specific Heat}

\begin{figure}
\begin{center}
\includegraphics[width=\linewidth]{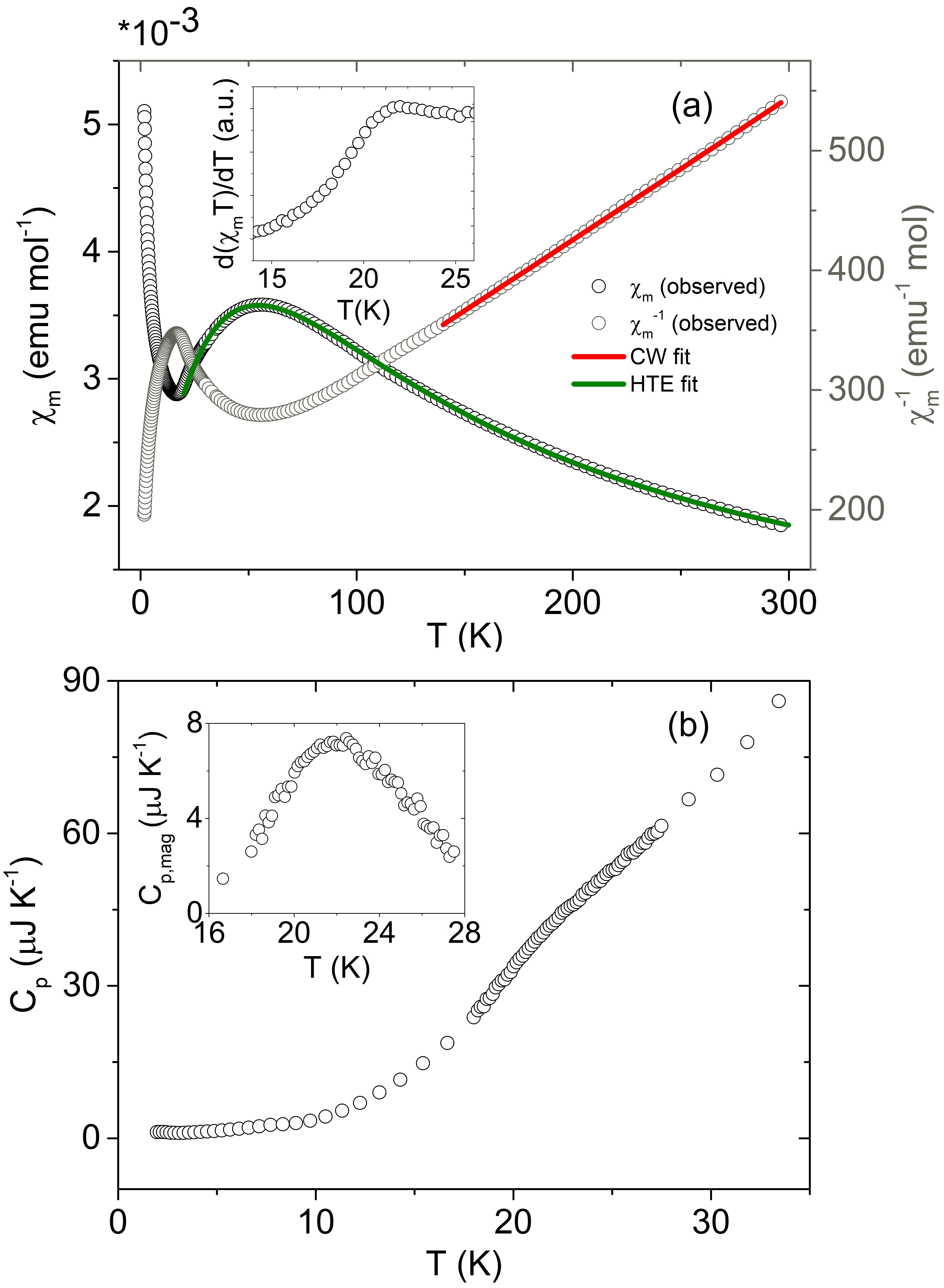}
%\hspace*{-9 cm} (a)
%\hspace*{+5.0cm} (b)
\end{center}
\caption{(a) Temperature dependence of the zero field cooled DC molar magnetic susceptibility (black circles) collected in a 1000 Oe applied magnetic field. The Curie-Weiss fit shown in red was obtained for a temperature interval of 200 K to 300 K and the extracted parameters are $C=0.88(2)$~emu~K~mol$^{-1}$, $\theta_{CW}=-167.0(5)$~K, and $\chi_0=-4.31(1) \times 10^{-5}$~emu~mol\textsuperscript{-1}. The green line represents the 10\textsuperscript{th} order high temperature series expansion fit using the $S=1$ anisotropic triangular lattice model with $J^\prime=0.28$~meV, $J=3.91(6)$~meV, and $\chi_0=4.86(6)\times 10^{-4}$~emu~mol\textsuperscript{-1}. Shown in the inset is the temperature derivative of the magnetic susceptibility as a function of temperature where a peaked feature is observed at $T_N=22$~K. (b) Temperature dependence of the specific heat $C_p$. The inset shows the magnetic contribution to the specific heat as obtained by subtracting the phonon contribution using a polynomial form $C_p=a_1T^3+a_2T^5+a_3T^7$ fitted to the high-temperature data. An anomaly, previously not observed in the previous work \cite{danis2018}, is seen at $T_N=22$~K.}\label{fig:thermodynamic}
\end{figure}

The temperature dependence of the molar magnetic susceptibility $\chi_m(T)$ of \bmp is shown in Fig.~\ref{fig:thermodynamic}(a). Within the temperature interval 200 to 300 K, $\chi_m(T)$ obeys the Curie-Weiss law $\chi_m(T)=C/(T-\theta)+\chi_0$, with a Curie constant $C=0.88(2)$~emu~K~mol$^{-1}$ and a large antiferromagnetic Weiss constant $\theta_{CW}=-167.0(5)$~K. Both values are consistently considerably higher than those reported in Hembacher \textit{et al.} \cite{danis2018} regardless of the temperature fitting range used. This could be relevant to the difference in sample quality and preparation method. The resulting effective moment $\mu_{eff} = 2.65(1) \mu_B$ ($g=1.94$) is close to the spin-only moment $2.83\mu_B$ expected for $S=1$. We also find a small $\chi_0=-4.31(1) \times 10^{-5}$~emu~mol\textsuperscript{-1}. 

Upon cooling, a broad maximum characteristic of short-range correlations is seen at $T_{max}= 55$~K, followed by an inflection point at $T_N=22$~K, associated with magnetic long range order. While the latter feature is consistent with the previous study \cite{danis2018}, we also find a previously unobserved anomaly in our specific heat data at this temperature [Fig.~\ref{fig:thermodynamic}(b) inset]. Both the presence of this anomaly and the smaller magnitude of the paramagnetic Curie tail in the low-temperature susceptibility are indicative of the high quality of our sample.

Given that previous first-principles calculations indicate that $J$ and $J^\prime$ should be the leading magnetic exchanges in \bmpns, and that the magnetic anisotropy should be small, we choose the $S=1$ Heisenberg anisotropic triangular lattice model as a starting point to fit the magnetic susceptibility. The \texttt{HTE10} code\cite{Lohmann} was used to calculate the $[4,6]$ Pad\'e approximant of the 10\textsuperscript{th} order high-temperature series expansion of this model. The calculated data were fitted to the observed $\chi_m(T)$ down to 40 K; $J$ and $\chi_0$ were allowed to vary while $J^\prime$ was fixed in $0.01J$ steps between $-0.5J$ to $J$. We find that the experimental $\chi_m(T)$ is best described [Fig.~\ref{fig:thermodynamic}] by the parameters $J=45.4(7)$~K ($3.91(6)$~meV), $J^\prime/J= -0.07$, a much reduced $g=1.44$, and $\chi_0=4.86(6)\times10^{-4}$~emu~mol$^{-1}$ \textit{i.e.} the interchain coupling is insignificant, and the system is effectively one-dimensional. We also note that the fitted $\chi_0$ is large and positive; this cannot be explained by the small amount of paramagnetic MoO$_2$ impurity present in the sample, but is plausibly accounted for by temperature-independent paramagnetism arising from the mixing in of higher-lying Mo$^{4+}$ orbital states. Indeed, the magnitude of $\chi_0$ is consistent with the temperature independent paramagnetism observed in other Mo$^{4+}$ containing compounds\cite{MoO2,TIP}. 

While the observed one-dimensionality is compatible with previous work on BaMo(PO$_4$)$_2$ \cite{danis2018} and other members of the anhydrous alum family, the relatively high ordering temperature of BaMo(PO$_4$)$_2$ remains unexplained, given that $J^\prime/J = -0.07$ is far away from the threshold value required to suppress the Haldane gap and induce magnetic order \footnote{Here, we assume that the fact that both ferro- and antiferromagnetic further neighbor couplings are frustrated means that the critical $J^\prime$ is similar for $J^\prime<0$ and $J^\prime>0$}. Furthermore, the ratio $\theta_{CW}/T_N$ is only $7.5$ in \bmpns, versus $>100$ for the isostructural $S=1/2$ material KTi(SO$_4$)$_2$ and $\sim 10$ for the $S=5/2$ system KFe(SO$_4$)$_2$, which also has much stronger interchain couplings. Both of these observations suggest the presence of other terms in the Hamiltonian as the cause of magnetic order in \bmpns.

\subsection{Magnetic Structure}
A comparison of the $1.5$~K and $30$~K~$>T_N$ long-wavelength NPD data taken on the WISH instrument reveals several magnetic Bragg peaks compatible with the commensurate propagation vector $\mathbf{k} = (\frac{1}{2}, \frac{1}{2}, \frac{1}{2})$ [Fig.~\ref{fig:HRPD}(b)]. To determine the magnetic space groups compatible with this propagation vector and the $C2/m$ nuclear space group, we used the \texttt{MAXMAGN} application on the Bilbao Crystallographic Server \cite{maxmagn}. This yielded a single magnetic space group, $P_S\overline{1}$ (in Belov-Neronova-Smirnova notation \cite{belov1957kristallografiyd}), with one Mo$^{4+}$ site and a free magnetic moment direction $(m_x,m_y,m_z)$. The magnetic model was fitted to the magnetic-only scattering obtained by subtracting $30$~K data from the $1.5$~K data; all structural and instrumental parameters were fixed to their nuclear structure refinement values at $30$~K, and only the three magnetic moment components were allowed to vary. To account for the strong hybridization between the Mo$^{4+}$ and O$^{2-}$ valence orbitals, we used a covalent magnetic form factor for Mo$^{4+}$ derived by Fourier transforming the spin density calculated from density functional theory \cite{Tsirlin}. The resulting fits are excellent [Fig.~\ref{fig:HRPD}(b)], with $R_{mag}=4.7\%$, and a slightly suppressed ordered moment of $1.37(5)\mu_B$ (versus the expected $gS\mu_B\sim1.44~\mu_B$ from the HTSE fit above) with a 44.33(6)\degree~ angle to the principal z axis. The individual moment components are ($0.96(5)$,$-0.04(1)$,$0.98(5)$). As illustrated in Fig.~\ref{fig:BMP_nuclear}, the magnetic structure is collinear and antiferromagnetic along the $b$ ($J$) direction and exhibits an antiferromagnetic stacking along $c$. Half the bonds along $\langle 110\rangle$ ($J^\prime$) are antiferromagnetic, and the other half ferromagnetic. Barring the insignificant component along the $b$-direction, this is the same magnetic structure as reported in \cite{danis2018}, with only a small difference in the angle between the $ab$-plane and the magnetic moment of $1.37(5)$ versus $1.42(9)$. Interestingly, the Bragg peaks from the present sample are resolution-limited, while those from the previous indicated a finite correlation length $\sim 50\AA$. Collinear magnetic order on the anisotropic triangular lattice is only stabilized for $J\gg J^\prime$ \cite{starykh}, which is in accordance with expectations from the magnetic susceptibility and \textit{ab-initio} calculations. On the other hand, it yields zero net exchange energy between the chains, regardless of whether the interchain coupling is antiferromagnetic or ferromagnetic. This, much like the fits to the magnetic susceptibility, implies a different magnetic ordering mechanism to interchain coupling in \bmpns.

\begin{figure}[h]
\begin{center}
\includegraphics[width=\linewidth]{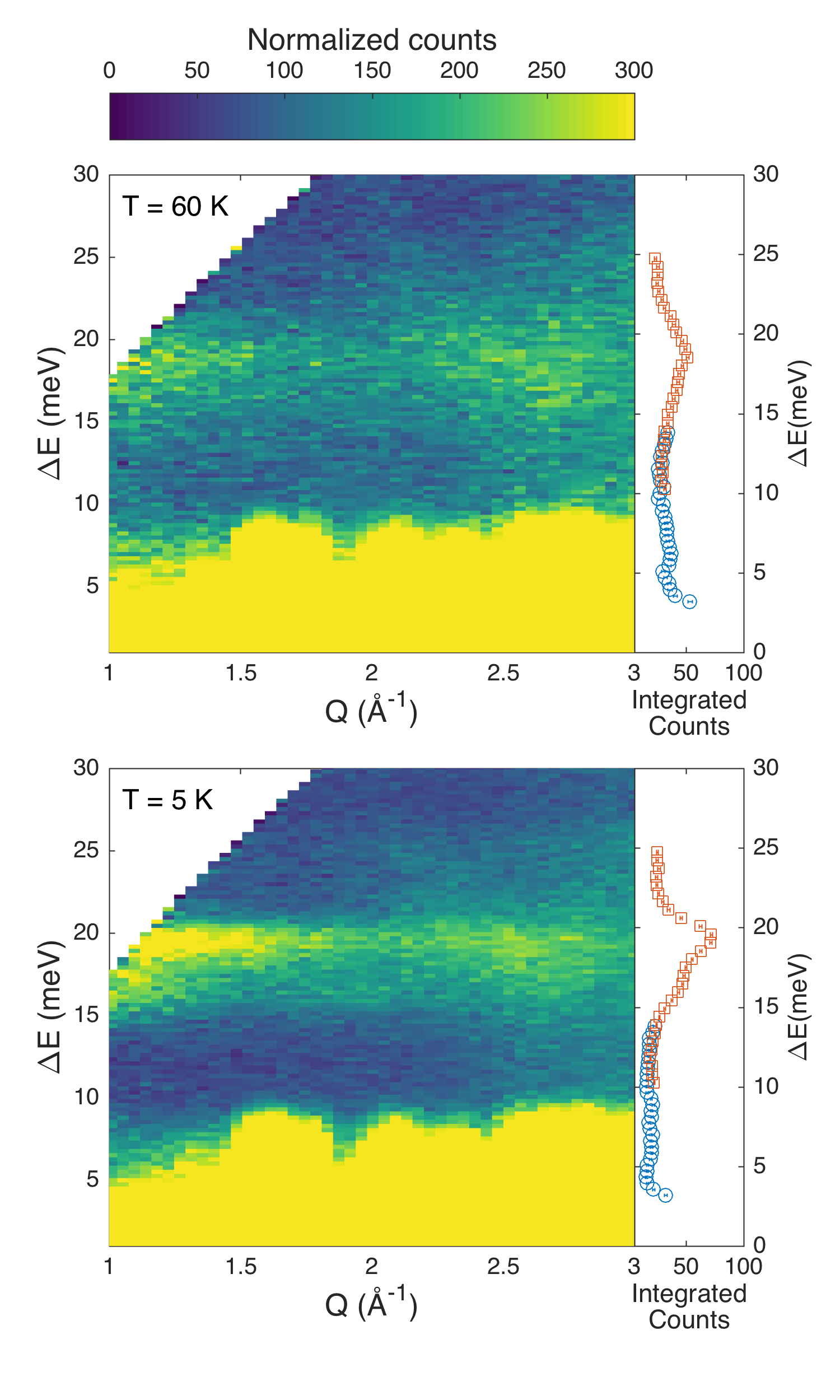}
%\hspace*{+1.3cm} (a)
\end{center}
\caption{$S(Q,\Delta E)$ at $T=60$~K (top) and $T=5$~K (bottom), along with energy integrated data for $E_i=50.4$~meV (red) and $E_i=19$~meV (blue)}\label{fig:INS}
\end{figure}

\begin{figure*}[t]
\begin{center}
\includegraphics[width=\linewidth]{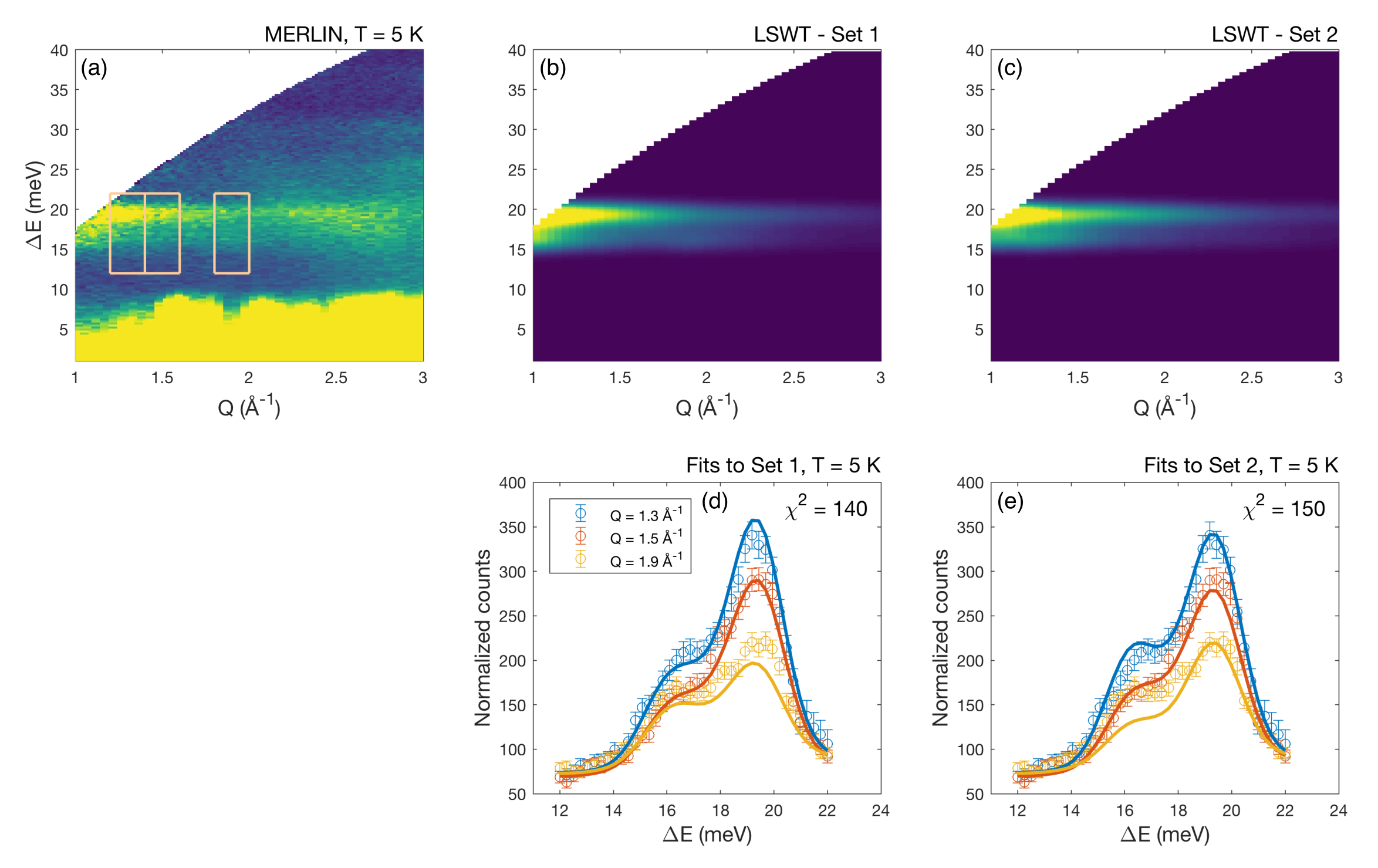}
%\hspace*{+1.3cm} (a)
\end{center}
\caption{(a) Experimental $S(Q,\Delta E)$ at 5~K, as in Figure 5. The orange boxes indicate the integration regions for the three cuts used in the fitting. (b-c) Simulated $S(Q, \Delta E)$ using parameter Sets 1 (b) and 2 (c) (Table.~\ref{Table:Model parameters}). (d-e) $\Delta E$-integrated cuts (open circles) fitted to parameter Set 1 (d) and Set 2 (e) (solid lines), as described in the text, for $Q=1.3$ (blue), $Q=1.5$ (red), and $Q=1.9$ (yellow).}\label{fig:LSWT}
\end{figure*}

\section{MAGNETIC EXCITATIONS AND HAMILTONIAN}
Although the magnetic susceptibility of \bmp is well fitted by an isotropic Heisenberg model with interchain coupling, this does not exclude a large easy-axis anisotropy as the ordering mechanism in \bmpns, particularly as the interchain coupling extracted from the fits is sub-critical. To answer this question, and to more generally elucidate the magnetic Hamiltonian of \bmpns, we carried out an inelastic neutron scattering experiment on the MERLIN spectrometer at the ISIS Neutron and Muon Facility. The dynamic structure factors $S(Q,\Delta E)$ collected at $E_i=50.4$~meV and at temperatures of $5$~K and $60$~K are shown in Fig.~\ref{fig:INS}. At $60$~K, well above $T_N$, the spectrum is gapless, but shows an accumulation of inelastic spectral weight in a broad peak centered at $18$~meV. On cooling below $T_N$, a gap develops in the spectrum, with the inelastic spectral weight now entirely concentrated in a narrow band between $14.5$~meV and $21$~meV. The presence of both a large gap and collinear magnetic order strongly suggest a large easy-axis anisotropy in the system; while a gapped excitation spectrum is consistent with both the large-$D$ and Haldane phases, neither of these can simultaneously show magnetic order. 

In order to extract the Hamiltonian parameters from the excitation spectrum in the ordered phase, we use the following anisotropic exchange Hamiltonian:

\begin{align}
\mathcal{H}=\sum_{i,j}\mathbf{S}_i\mathcal{J}^{\mu\nu}\mathbf{S}_j+J^{\prime}\sum_{i,j}\mathbf{S}_i\cdot\mathbf{S}_j+D\sum_i{(\mathbf{S}_i^z)^2}
\label{Magnetic_model}
\end{align}
\noindent where $\mathcal{J}^{\mu\nu}$ is the exchange tensor between intra-chain neighboring spins along the $b$ direction, $J^\prime$ is the isotropic inter-chain neighbor exchange along $\langle 110 \rangle$, and $D$ is an effective axial anisotropy, which replaces the full anisotropy tensor expected for the $2/m$ site symmetry (this choice is justified by our later DFT calculations). The indices $\mu,\nu=\{x,y,z\}$ in the exchange tensor refer to a Cartesian basis where $y||b$, and $x$ and $z$ are rotated about the $b-$axis so that $z$ coincides with the refined easy axis from our NPD data. The inversion center at the midpoint of the bonds between the nearest-neighbor Mo$^{4+}$ ions eliminates the anti-symmetric part of the exchange tensor, leaving the following allowed elements:
\begin{align}
\mathcal{J}^{\mu\nu}=\begin{pmatrix}
J^{xx} & 0 & J^{xz} \\
0 & J^{yy} & 0\\
J^{xz} & 0 & J^{zz}
\end{pmatrix}
\end{align}

The experimental data were fitted to the above model using the \texttt{SpinW} software \cite{spinw}, which calculates the dynamic structure factor of spin systems using linear spin-wave theory \footnote{Linear spin wave theory ignores the quantum corrections to the dispersion bandwidth and gap expected for low spin systems, and the parameters extracted should therefore be taken as indicative.}. Three cuts through the experimental $S(Q,\Delta E)$ along $\Delta E$ were chosen to optimize the model against; these were centered at $Q={1.3,~1.5,~1.9}$~\AA$^{-1}$ and integrated over $dQ=0.2$~\AA$^{-1}$, as shown in Fig.~\ref{fig:LSWT}. The fitting was carried out using a particle swarm optimization algorithm, where the powder average of the linear spin-wave intensities were calculated and convoluted with a Gaussian resolution at each step. After several runs of the algorithm, two solutions with similar goodness-of-fit $\chi^2$ were found, which will be referred to as Set 1 and Set 2 for the remainder of the paper; the parameters for each are listed in Table \ref{model_params}. The error bars on the parameters was estimated from the standard deviation of the final positions of the particles with the lowest $\chi^2$. 

While Set 1 corresponds to a moderate easy-axis exchange anisotropy along the same direction as the single-ion anisotropy, Set 2 shows an easy-plane exchange anisotropy, with the plane defined by the magnetic moment direction and the $b$ axis. The single-ion terms are similarly large in both cases, with $D/J^{zz}\sim -1.5$ and $-1.6$, respectively, thus exceeding estimates from previous \textit{ab-initio} calculations by more than an order of magnitude. On the other hand, the small values of the inter-chain coupling, $J^\prime/J^{zz}=0.08$ for Set 1 and $0.05$ for Set 2, are consistent with these calculations. While the goodness-of-fit for the neutron data slightly favors Set 1 ($\chi^2=140$) over Set 2 ($\chi^2=150$), the two are difficult to distinguish from the spin-wave fits alone, especially because scattering from acoustic phonons contaminate the magnetic signal from $Q\sim2$~\AA~ and outwards. A definite determination of the magnetic Hamiltonian must therefore await the availability of single crystals.

Before concluding the section, we note that fits to several simpler models were attempted before turning to the anisotropic model Hamiltonian above: the best of these was for a model with isotropic $J$ and $J^\prime$ and a single-ion anisotropy $D$, which resulted in $\chi^2=165$. However, this model yielded an unreasonably large $J^\prime/J=0.7$. We also note that it was not possible to check the consistency of the anisotropic model with the magnetic susceptibility because no publicly available codes are able to calculate the model with anisotropic exchange, single ion anisotropy, and frustrated interchain coupling simultaneously. The apparent agreement between the small $J^\prime$ extracted from the present spin-wave fits and those to the susceptibility above may therefore be accidental.

\section{Density Functional Theory Calculations}
\label{sec:methods}

Our previous calculations showed only a weak single-ion anisotropy $D<0.1$~meV \footnote{In Ref.~\onlinecite{danis2018}, we mistakenly introduced a pre-factor of $\frac18$ when calculating total energies per Mo atom.}, in clear contradiction with the large $D$ value extracted from the linear spin-wave fits above. In addition, the spin direction determined by the minimum of the total energy did not match the experimental spin direction from neutron diffraction, although the latter was found to coincide with the direction of the maximum orbital moment~\cite{danis2018}.

In order to resolve these discrepancies, the parameters of the spin Hamiltonian (Eq.~\ref{Magnetic_model}) were re-determined on the level of density functional theory (DFT) by calculations within the generalized gradient approximation (GGA)~\cite{pbe96} using both the Projector Augmented Wave (PAW) based Vienna \textit{ab-initio} Simulation Package (\texttt{VASP})~\cite{vasp1,vasp2} with the 700\,eV plane-wave energy cutoff, as in Ref.~\cite{danis2018}, and the full-potential linearized augmented planewave (LAPW) based \texttt{ELK}~\cite{ELK} code with $R_{MT}*|G+k|_{max}$=7.5. 

All calculations were performed on the 8$\times$8$\times$8 $\mathbf{k}$-mesh centered at the $\Gamma$ point with an energy convergence criteria of 10$^{-6}$\,eV/f.u. Increasing the $\mathbf k$-mesh to $10\times 10\times 10$ changed individual anisotropy energies by less than $10^{-3}$\,meV, suggesting good convergence with respect to the number of $\mathbf k$-points. Spin-orbit coupling effects were included self-consistently in both methods, and the experimental structural parameters were used without relaxation. Strong correlations in the Mo $4d$ shell were taken into account within DFT+$U$, where we set the Coulomb repulsion to $U=3$\,eV, as obtained by the linear-response method~\cite{cococcioni2005} in the previous study~\cite{danis2018}. Simultaneously, we noticed that $J_H$, the on-site Hund's coupling within DFT+$U$, has a strong influence on the magnetic anisotropy of the system. Thus, we adopted several values of $J_H$ for the calculations, which will be discussed below. The fully localized form of the double-counting correction was employed~\cite{FLL-DC}. 

\begin{table}
\caption{\label{model_params}Model parameters for spin Hamiltonian (Eq.~\ref{Magnetic_model}).}
\begin{ruledtabular}
\hspace{-0.75cm}
\begin{tabular}{@{}rcccc}
%\br
~ &  $\mathcal{J}^{\mu\nu}$ (meV) & $D$ (meV) & $\theta_D$ (\degree) & $J^\prime/J^{zz}$ \\
\hline \\
Set 1 & $\begin{pmatrix}
3.3(1) & 0 &  -0.2(1) \\
0 & 3.34(6) & 0\\
-0.2(1) & 0 & 4.1(1)
\end{pmatrix}$ \vspace{0.5cm} & -6.01(8) & $0$ & 0.08 \\
Set 2 & $\begin{pmatrix}
3.2(1) & 0 &  -0.1(1) \\
0 & 4.03(5) & 0\\
-0.1(1) & 0 & 4.0(1)
\end{pmatrix}$ \vspace{0.5cm} & -6.30(9) & $0$  & 0.05 \\
\texttt{VASP} & $\begin{pmatrix}
3.00 & 0 &  -0.60 \\
0 & 5.98 & 0\\
-0.60 & 0 & 4.95
\end{pmatrix}$ \vspace{0.5cm} & -2.50 & 6 & 0.08 \\
\texttt{ELK} & $\begin{pmatrix}
5.20 & 0 &  -0.25 \\
0 & 5.58 & 0\\
-0.25 & 0 & 5.32
\end{pmatrix}$ \vspace{0.5cm} & -2.65 & 8 & 0.07 \\
\end{tabular}
\end{ruledtabular}
\label{Table:Model parameters}
\end{table}

\begin{figure}
\includegraphics[width=0.49\textwidth]{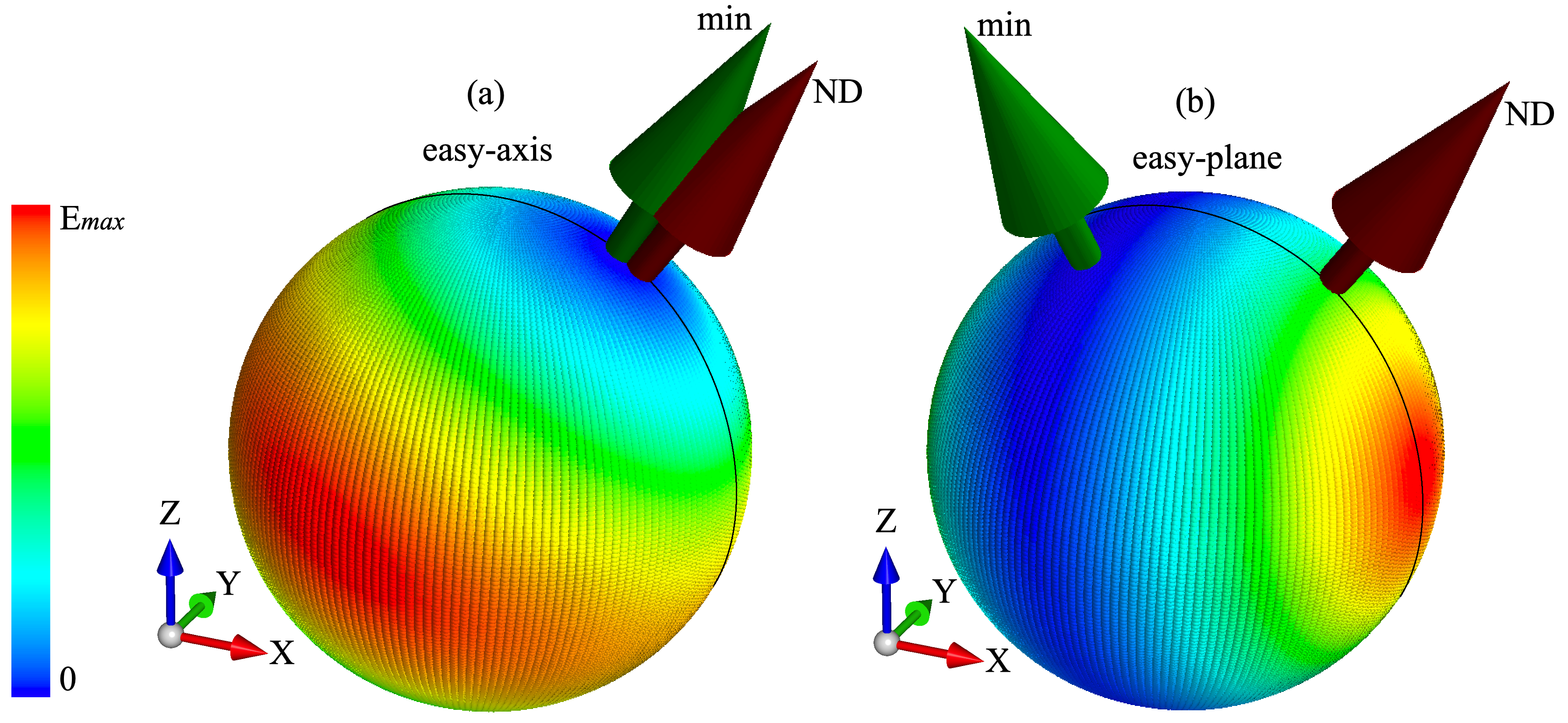}
\caption{3D map of the magnetic anisotropy energy obtained by interpolating a finite number of configurations with different directions of the magnetic moment within the primitive unit cell (one Mo atom) and using the PAW-based \texttt{VASP} code. Panels (a) and (b) show the results for $J_H=0.6$ eV (E$_{max}$ = 3.76 meV) and $J_H=0.8$ eV (E$_{max}$ = 2.04 meV), respectively.  The black line is the rotation plane used in Fig.~\ref{fig:SIA}. The green arrow labeled "min" is the direction of the energy minimum, whereas ND stands for the experimental spin direction determined from neutron diffraction. A switch between an effective easy-axis to an effective easy plane anisotropy is observed on increasing $J_H$.}
\label{fig:3D_anisotropy}
\end{figure}

First, the dependence of the PAW-based \texttt{VASP} results on the DFT+$U$ parameters was investigated. Changing the value of $U$ within the physically reasonable range of 1 -- 2\,eV had no significant influence on the computed anisotropy. On the other hand, the anisotropy is remarkably sensitive to the on-site Hund's coupling $J_H$. By reducing $J_H$ from 0.8\,eV used in Ref.~\onlinecite{danis2018} to 0.6\,eV, the experimental easy-axis scenario can be reproduced. On the other hand, at $J_H\gtrsim $ 0.7\,eV an easy-plane anisotropy is obtained, at odds with the experimental results [Fig.~\ref{fig:3D_anisotropy}]. A similar evolution of the anisotropy with $J_H$ was previously observed in Sr$_3$NiPtO$_6$\cite{Pradipto2016} albeit with a higher threshold value lying above 1\,eV for the $3d$ Ni$^{2+}$ ion.

Since the magnetic anisotropy amounts to a tiny energy difference between calculations for different parameters, numerical effects become important. Therefore, we cross-checked our \texttt{VASP} results using the full-potential \texttt{ELK} code using a different basis set. The strong dependence of the anisotropy on $J_H$ was reproduced, although in this case the transformation from the easy-axis scenario to the easy-plane one took place around $J_H=0.9$ eV instead of 0.7 eV in \texttt{VASP}. This difference can be attributed to the different basis sets and, consequently, different orbital occupations that enter the energy correction of DFT+$U$. For a given set of $U$ and $J_H$ values, the spin-diagonal components of the occupation matrices differ by more than 20\% between the codes, thus resulting in very different total numbers of Mo-$4d$ electrons, 4.07 and 3.24 in VASP and ELK, respectively. Since the energy correction of GGA+$U$ explicitly depends on the orbital occupations, the same values of the Hubbard $U$ and Hund's $J_H$ may therefore lead to different results in the two codes. 

\begin{figure}
\includegraphics[width=0.49\textwidth]{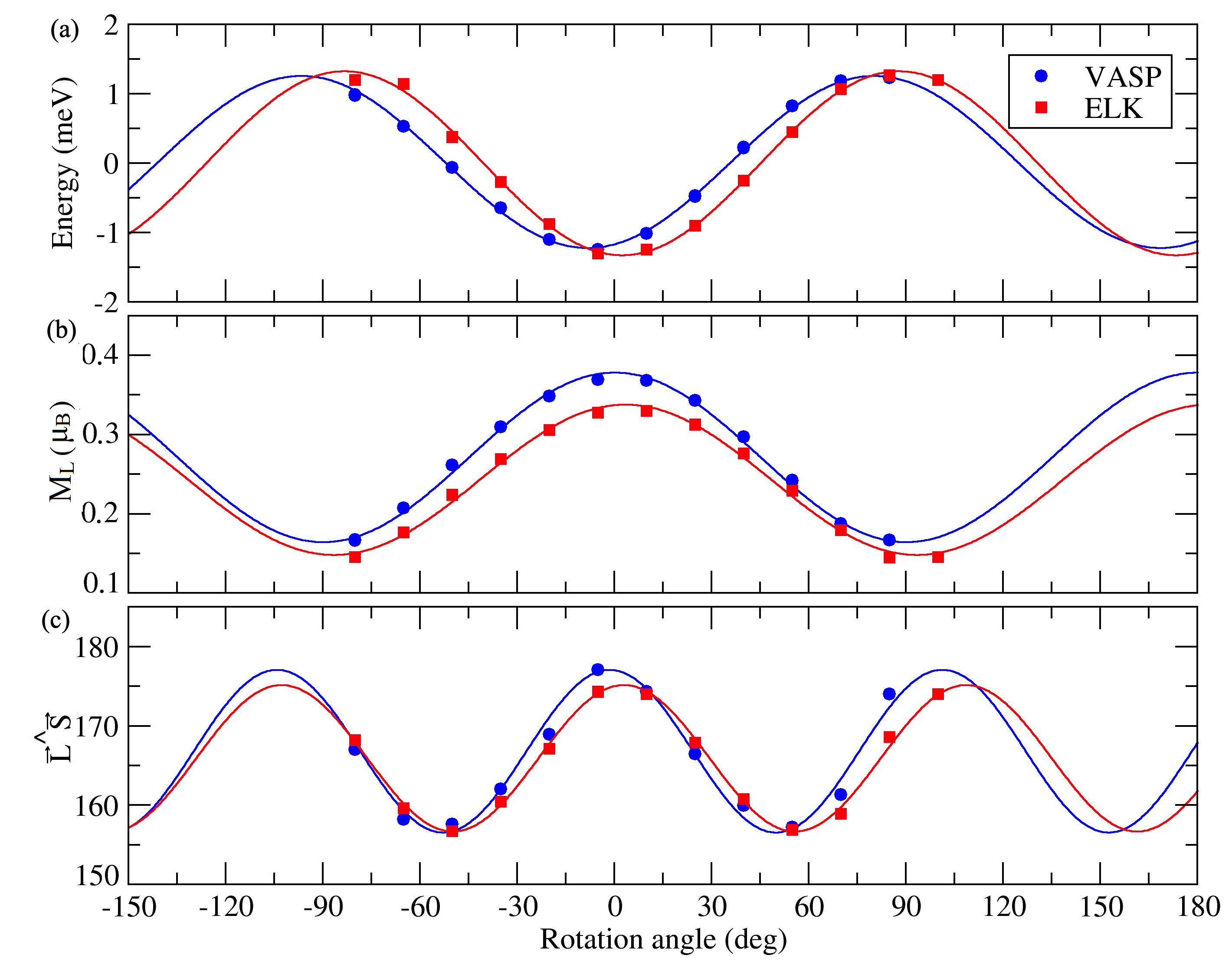}
\caption{(a) Anisotropy energy, (b) angular dependence of the orbital moment, and (c) the angle between the spin and orbital moments depending on the spin direction in the $xz$-plane. The experimental spin direction $\boldsymbol{\gamma} \sim (0.70, 0, 0.72)$ is chosen as zero. The results are obtained using $J_H$=0.6 eV. Note that the spin moment $M_S$ = 2 $\mu_B$ stays constant within this rotation.}
\label{fig:SIA}
\end{figure}

Remarkably, the change in the character of the anisotropy with changing $J_H$ reflects not only the transformation between the easy-axis and easy-plane scenarios, but also between the different microscopic mechanisms behind them. Conventional theories of magnetocrystalline anisotropy (Bruno's model)~\cite{bloch1931, bruno1989, Bruno_rule} suggest that the orbital moment is maximized when the magnetization points along the easy axis. This is true for the easy-axis scenario obtained at low $J_H$ [Fig.~\ref{fig:SIA}], but no longer holds for the easy-plane scenario at high $J_H$, because the direction of the maximum orbital moment does not change with $J_H$, and always coincides with the experimental easy axis. As such, Mo$^{4+}$ follows the conventional mechanism for small $J_H$, whereas it does not for larger $J_H$.

The correlation between the easy axis and orbital moment implied by Bruno's model stems from the perturbative treatment of the spin-orbit coupling, where spin-flip transitions are neglected, and the spin-majority electrons do not contribute to the orbital moment~\cite{Laan_2001}. It is thus conceivable that such processes are increasingly important at higher $J_H$, and the simple perturbative treatment becomes invalid. However, experimentally one may not reach this regime, because in $4d$ metals lower values of $J_H=0.4-0.6$ eV are generally assumed~\cite{Okamoto2017, Mravlje2012, Iqbal2017}. 

We therefore choose $J_H$ = 0.6\,eV for use in further calculations. In Fig.~\ref{fig:3D_anisotropy}(a), we show the polar plot of the magnetic energies, which represents the joint effect of the single-ion and exchange anisotropies. Compared to the previous calculations\cite{danis2018}, one observes a good agreement between the calculated energy minimum and experimental easy-axis direction from powder neutron diffraction. This way, we conclude that the earlier \textit{ab initio} determination of the anisotropy was compromised by the large $J_H$ value. A realistic $J_H$ leads to an easy-axis anisotropy scenario, in good agreement with experiment.

To gain further insight into the anisotropy components, we discriminate between the on-site and exchange contributions using a magnetic supercell and the mapping procedure described in Ref.~\cite{xiang2011}. The resulting single-ion anisotropy and exchange tensors are listed in Table \ref{model_params} and show a qualitative agreement with Set 2 from the linear spin-wave fits. More specifically, we obtain an exchange tensor where the $yy$ and $zz$ components largely exceed the $xx$ one. The remaining discrepancies may be caused by the choice of the $U$ parameter on the DFT side and by not accounting for quantum effects in the linear spin-wave theory fits on the experimental side. Moreover, DFT is known to underestimate the single-ion anisotropy in other transition-metal compounds~\cite{Pradipto2016}.

The magnetic moments from DFT are around 2 $\mu_B$ for the spin part and $-0.35$ $\mu_B$ for the orbital part. The total moment of 1.65 $\mu_B$ slightly exceeds the experimental value of 1.37(5) $\mu_B$, but this is not unexpected, given quantum fluctuations associated with the low spin and dimensionality. Lastly, we note that even if DFT predicts a lower $D/J\sim -0.54$ compared to $\sim -1.5$ from experiment, this is still sufficient to close the Haldane gap and trigger long-range magnetic order in BaMo(PO$_4)_2$.

\section{Discussion}
Both the fits to our inelastic neutron scattering data and our electronic structure calculations indicate a strong single-ion anisotropy and a weaker exchange anisotropy in \bmpns. The former also accounts for the relatively large ordered moment observed in neutron diffraction; series expansion calculations predict that the Haldane phase is suppressed at a critical $D/J=-0.295$ \cite{Rommelse}, with a sharp increase in the ordered moment from around $60\%$ to $90\%$ of the full ordered moment $gS=2\mu_B$ upon increasing $D/J$ to $-1$. The experimental ordered moment extracted from neutron diffraction is $70\%$ of the saturation value estimated from the Curie-Weiss fit; this value is considerably in excess of what is expected for a system where the order is induced by interchain couplings. The same is true of the high ordering temperature, and the fact that the order is collinear, rather than a coplanar helix, as is predicted for the cases of no or easy-plane anisotropy.

Having conclusively demonstrated the strong easy-axis anisotropy in \bmpns, we now turn to its implications on the low-temperature physics of the material. The $S=1$ chain model with easy-axis anisotropy is expected to manifest excitation continua in its dynamics, despite having a long-ranged ordered ground state; these continua are often described in terms of an effective $S=1/2$ model, where the $S^z=\pm 1$ states correspond to $S^z_{eff}=\pm\frac12$, and $S^z=0$ to holes along the chain. The lowest-lying excitations in this picture are states with a single hole; this hole can either delocalize, creating fractional ``holon'' and ``spinon'' quasiparticles, or the excitation can be accompanied by a flip of a domain on either side of the hole, creating a single holon \footnote{The creation of a single hole results on in a spin configuration along the chains of $...-+-0-+-...$ (or the equivalent with all spins reversed), where $+$ and $-$ represent the $S_{eff}=\pm \frac12$ states. When the hole hops along the chain it will leave either a $S^z_{eff}=+\frac{1}{2}$ or $S^z_{eff}=-\frac{1}{2}$ spin at its previous site. If this is parallel to the neighboring spin (e.g. the hop transforms $...-+-0-+-...$ into $...-+--0+-...$), as is the case when exciting from the N\'{e}el ground state, a domain wall (here $--$) has been created at energy cost $J$; by analogy with the $S=\frac{1}{2}$ Ising chain, this domain wall is a ``spinon'', while while the mobile hole is called a ``holon''. In the other case, \textit{i.e.} if the antiferromagnetic domain on either side of the holon is flipped when the hole is created, there is no energy cost and only a delocalized holon\cite{Rommelse,denNijs}.}. It is the latter of these which is lowest in energy, and the condensation of single-holon excitations is associated with the transition between the N\'{e}el and Haldane states. In the conventional language of magnons, the fractionalized continua of the effective $S=1/2$ model correspond to multi-magnon processes, with even numbers of magnons appearing in the transverse channel and odd numbers in the longitudinal. Recent matrix product calculations in the same region of the phase diagram as \bmp indicate that these high-order multi-magnon processes are expected to be particularly intense in the longitudinal channel at elevated temperature and at low energy \cite{Lange}. As far as we know, such continua have not yet been observed experimentally.

\section{Conclusion}
By combining experimental results from bulk measurements and elastic and inelastic neutron scattering with first principle electronic structural calculations, we have re-evaluated the magnetic Hamiltonian and ordering mechanism of the anhydrous alum \bmpns. The analysis of NPD data suggest collinear antiferromagnetic order at $T_N=22$~K, similar to a previous study. On the other hand, the sample used in the present study exhibits resolution limited magnetic Bragg peaks, a thus far unseen peak at $T_N$ in the specific heat, and a larger Curie-Weiss constant $\theta$ than previously reported, indicating that our new synthetic procedure yields significantly higher quality samples of \bmpns than before. Since high-temperature series expansion fits to $\chi_m(T)$ indicate that the inter-chain coupling cannot alone stabilize the magnetic order, we searched for anisotropic terms in the magnetic Hamiltonian using inelastic neutron scattering measurements. These revealed a considerable easy-axis single-ion anisotropy, which was estimated to be between $D \sim -6.3$~meV and $-6.0$~meV \textit{via} linear spin-wave theory fits to the experimental dynamic structure factor $S(Q,\Delta E)$. The fits also yielded a small $J^\prime = 0.05-0.08J$. Our revised first principles calculations, which account for the strong dependence of the anisotropy on the Hund's coupling $J_H$, are able to qualitatively reproduce all of these parameters, as well as the ordered moment direction. Our study therefore clarifies the ordering mechanism and the ground state Hamiltonian of \bmpns. The Hamiltonian parameters extracted from the spin wave fits and electronic structure calculations place BaMo(PO\textsubscript{4})\textsubscript{2} within the rarely accessed region of the anisotropic $S=1$ chain phase diagram where a dominant easy-axis anisotropy causes both magnetic order and exotic excitation continua. The latter will be investigated in future neutron and X-ray scattering experiments on single crystal samples. 

\section{Acknowledgements}
The work of D.I.B. was funded by RFBR according to the research project No. 18-32-00018. A.A.T. acknowledges financial support by the Federal Ministry for Education and Research through the Sofja Kovalevskaya Award of Alexander von Humboldt Foundation. D.I.B. and Y.O.K. acknowledge the support by The Swedish Foundation for International Cooperation in Research and Higher Education (STINT).  The work at Chalmers University of Technology was supported by MAX4ESS under project number CTH-012. Financial support for the PhD of A.H.A by the University of Liverpool and the Science and Technology Facilities Council (STFC) is acknowledged. We gratefully acknowledge the STFC for access to neutron beamtime at ISIS and thank Dr. Gavin Stenning for aiding with SQUID and specific heat measurements at the Materials Characterization Laboratory, ISIS.  We thank Florian Lange (Uni. Greifswald), Holger Fehske (Uni. Greifswald), Lucy Clark (Uni. Liverpool), and Frank Kruger (ISIS and University College London) for useful discussions.

\bibliography{bibliography}{}
%\printbibliography

\end{document}